# AN ICT-BASED REAL-TIME SURVEILLANCE SYSTEM FOR CONTROLLING DENGUE EPIDEMIC IN SRI LANKA

A. Rukshan [1], A. Miroshan [2]

[1] Faculty of Business Studies, Vavuniya Campus, University of Jaffna, Sri Lanka

Email: rukshan@mail.vau.jfn.ac.lk

[2] Agriculture Division, Divisional Secretariat, Kyts, Jaffna, Sri Lanka

Email: miroshan_alexander@yahoo.com

## ABSTRACT

*Dengue is a notifiable communicable disease in Sri Lanka since 1996. Dengue fever spread rapidly among people living in most of the districts of Sri Lanka. The present notification system of dengue communicable diseases which is enforced by law is a passive surveillance system carried out by the public health care professionals. The present notification of communicable disease system is manual, slow, inefficient, and repetitive all of these lead to handle the dengue related health problems ineffectively. Thus it is less effective in preventing a spreading epidemic, public health care professionals and others require an operational support system to help for managing day-to-day public health responsibilities as well as a method to effectively detect and manage health problems such as Dengue. On the other hand the Information and Communication Technology (ICT) in medical world has been widely used. To give the information technology touch, a complementary web based open source software application environment has been developed with minimum implementation and recurrent costs critical for developing countries like Sri Lanka and named as eDCS: e Dengue Control System based on the same principles of manual disease surveillance system while taking steps to provide timely, accurate information in a reliable and useable manner. The eDCS helps to manage outbreaks through early detection, rapid verification, and appropriate response to Dengue. It allows health care professionals and citizens to get early awareness about the dengue disease via Internet or mobile phone and bring them for performing Dengue prevention and controlling operation through the social media acceleration. The system is initially limited to dengue communicable disease. It can be easily expanded to other communicable diseases, and non communicable disease surveillance in future.*

## 1. INTRODUCTION

Sri Lanka is classified as a "Category A" country by World Health Organisation (WHO), which means dengue fever is a leading cause of hospitalization and death among children; there are cyclical epidemics in urban areas; and the virus is a major public health concern. The disease has become a huge threat to health of Sri Lankans' and to the social-economic stability as well. During the last 12 months of the year 2012, 44456 dengue cases have been reported to the Epidemiology Unit from all over the island (Epidemiology Unit, 2013). Situation is same for many other South Asian countries as well. The only way to get rid of this danger is the extinction of the cause-mosquitoes by properly identifies its breeding sites or place (Rupasinghe et al, 2010). Public health care professionals agree that identifying and immediately removing dengue mosquito breeding places or sites is the most important step in mitigating risk of the dengue epidemic (DailyFT, 2012). Disease prevention and control measures have been established for early detection and monitoring of outbreaks. Unfortunately, the lack of centralized data and organized resources and capital in some countries like Sri Lanka has resulted in a number of increasing dengue outbreak cases (Matthews et al., 2009). The aim of this research is to have the ICT as one of the Sri Lankan tools to centralise data, and integrating existing web services for controlling dengue in the country. To give the information technology touch, a complementary web based open source environment software application has been developed and named as eDCS: **e Dengue Control System** based on the same principles of paper based disease surveillance system while taking steps to improve the quality and timeliness of data. Currently the data collection, information analyses, early outbreak detection, and decision making has a huge time gap. Our project objectives are to reduce this time gap within one to two days and thereby increasing the ability to identify disease outbreaks rapidly and increase the ability to take proactive measures thereby increasing agility in controlling diseases outbreaks.

## 2. METHODS

### 2.1 Problems in the Current Dengue Notification System

The communicable disease notification system is one of the most significant pillars of the preventive health care system in Sri Lanka. The current notification system which is enforced by law (The Quarantine and Prevention of Diseases Ordinance Gazette No: 7481 of 1925 August 28) is a passive surveillance system. The flow of the information and process of the present system is shown in Figure 1.

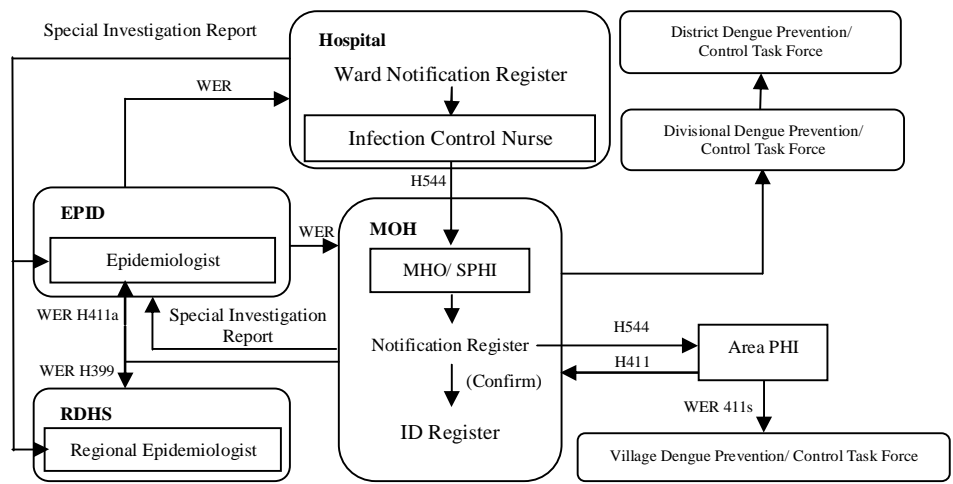

**Figure 1: The present Dengue Notification System**

where:
MHO: Medical Health Officer; SPHI: Supervising Public Health Inspector; PHI: Public Health Inspector;
MOH: Medical Officers of Health; RDHS: Regional Directorate of Health Services; EPID: Epidemiology Unit of Sri Lanka;
ID: Infection Disease; WER: Weekly Reporting
H399: Weekly Return of Communicable Diseases, Form designed by Ministry of Health, Sri Lanka
H411: Communicable Disease Report - Part I, Forms designed by Ministry of Health, Sri Lanka
H411a: Communicable Disease Report - Part II, Forms designed by Ministry of Health, Sri Lanka
H411s: Weekly Reporting Forms designed by Ministry of Health, Sri Lanka
H544: Notification of Communicable Disease, Forms designed by Ministry of Health, Sri Lanka

| S.No | Flow of Notification | Communication Method | Approximate Time Taken (Days) |
|------|---------------------|---------------------|------|
| 1 | Hospital to MOH office | Postal Service | 6 |
| 2 | MOH office to PHI | Official visit to MOH office | 2 |
| 3 | PHI visits the patient home | Official visit | 2 |
| 4 | PHI reports back to MOH | Official visit to MOH office | 2 |
| The total number of days taken for processing an outbreak. | | | 12 |

**Table 1: The present information flow of the dengue notification system in Sri Lanka.**

The patient with a Dengue communicable disease is admitted to the hospital the approximate time take to data processing, communication, and information sharing among various entities for notifying would be as given in Table 1. According to the current system dengue disease data are collected inpatients of public hospitals. When a patient with a Dengue Fever (DF)/ Dengue Hemorrhagic Fever (DHF) disease is seen by a doctor in a hospital, then a *Notification of Communicable Disease* form (H544) is filled by the Infection Control Nurse (ICN) and sent to the Medical Officers of Health (MOH) office where the patient is residing. The notification is sent on suspicion of the disease. On receiving the notification form by the MOH office, the Public Health Inspector (PHI) are assigned by MOH officials to investigate the cases related to their regions for confirmation where the patient is residing. The PHI visits the patient residence and takes investigation and measures for DF/DHF disease. The investigation results are sent back to the MOH offices. Once the disease is confirmed it is entered in the Infectious Disease (ID) Register of the MOH Office. Further, the PHI ensures that the patient is taking proper treatment and encourages continued treatment. The PHI observes the environment to locate the potential source of infection and take prevention and control measures to avoid further spread in the community. One step further, the PHI provides necessary health education to relevant people in the region and takes measures to prevent the future outbreaks.

The MOH offices prepare the form of *Weekly return of communicable diseases* (H399) containing the summary of the reported cases during a particular week incorporating the confirmation Dengue disease data. This summary is sent weekly as post to the Epidemiology Unit (EU) and to the Regional Director of Health Services.

The present dengue notification system is a fully paper-based and communication by post which contain the following weakness which are lack of active surveillance, unsatisfactory of timeliness, lack of laboratory surveillance, lack of coordination between front-line health care professionals and technical experts, lack of data consistency, lack of contribution of private sector, lack of case management, lack of integrating ICT resources and services, and lack of ICT based public awareness and community engagement. It is clear that the present notification for the Dengue communicable disease surveillance is a manual system which is slow in time (takes approximately 12 days to complete the whole cycle), provide poor quality of data in terms of readability and accuracy, less efficient. Figure 2 illustrates the possible situation of dengue outbreak due to delayed response. Public Health professionals and others require an operational support system to help for managing day-to-day public health responsibilities as well as a method to effectively detect and manage health problems such as dengue, and other communicable diseases, all of which may pose an outbreak threat in Sri Lanka. It is obvious that Dengue communicable disease notification system should provide information as rapid in time, and perfect in a reliable.

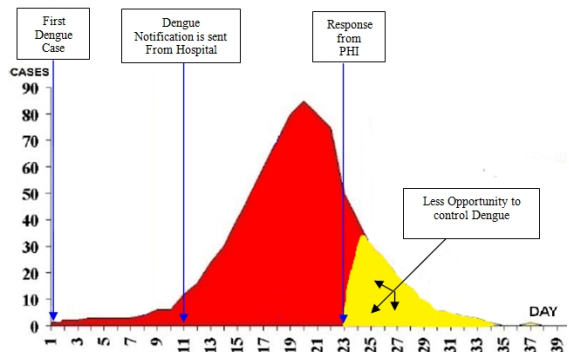

**Figure 2: Outcome of delayed response to dengue outbreaks**

## 2.2 Our Approach

The authors propose a solution through this research to overcome the above mentioned problems by adopting two major components such as research component and practical development component.

In practical part of our research, we propose a system of eDCS: **eD**engue **C**ontrol **S**ystem which collects data as early as possible and converting information to early action for preventing and controlling Dengue by integrating ICT and existing web services in Sri Lankan prospective. Furthermore, the eDCS provides public health professionals with integrated tools that assist in collecting data, analysing, monitoring, managing, and reporting on public health. The eDCS provides both front-line service providers and public health decision makers with critical information and reuses centralized data where possible. One step further, the eDCS provides information for the responsible citizens to access it for the prevention and control measures. Our approach is cost effective and can improve the quality of life in developing countries like Sri Lanka by accurately detecting dengue and providing data to health care professionals for monitoring dengue.

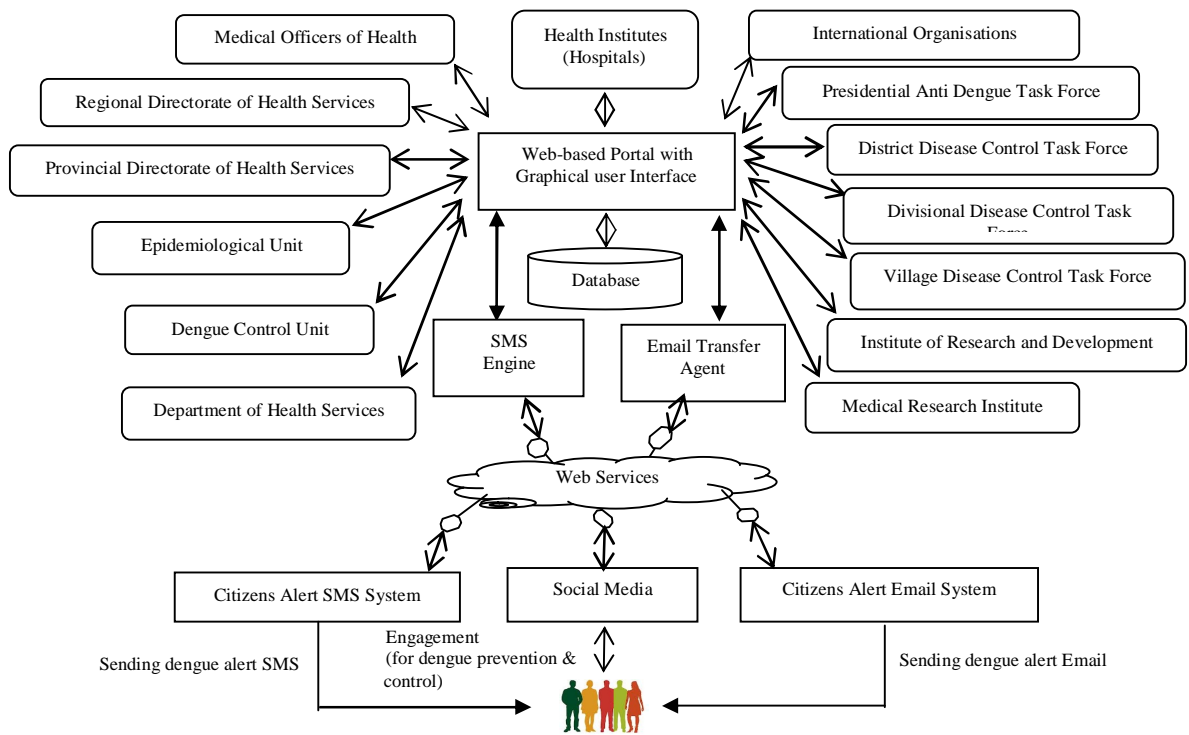

**Figure 3: The eDCS architecture**

The system has been constructed in order to fulfil the following goals: Provide accurate, usable, timely information to prevent and control the spread of Dengue communicable disease; Identify the correct source and place from where the Dengue spread; Provide a summarised and detail live Dengue map of Sri Lankan; Integrate the system with other ICT resources and web services where the API is available to prevent, control, analyse, and research the disease; In cooperate the public and private health sector to eliminate the disease from the country.

## 3. SYSTEM ARCHITECTURE

In order to reach the goals stated, our system activities can be characterized by the following sequence of activities;

1. Store Dengue infected patient basic information.
2. Compute and visualize the information of identified new cases.
3. Store patient travel history (for last 14 days) information.
4. Compute the information of identified risk places and visualize by considering patient travel history information.
5. Compute Dengue disease real-time report.

We critically review several related research papers and articles from Epi Info (2012), Cakici (2011), CDC (2011), HealthMap (2011), Kalpana (2011), Freifeld *et al*. (2010), Kass-Hout (2009), RODS (2009), Freifeld *et al*. (2008), Rolfhamre *et al*. (2006), Bradley *et al*. (2005), Wurtz (2005), Espino *et al*. (2004), Hamby *et al*. (2004) Voetsch (2004), Tsui *et al*. (2003), and Effler *et al*. (1999) for proposing our approach. The system architecture is as given in Figure 3.

## 4. RESULT

The web based software application has been given in Figure 4, Figure 5, and Figure 6.

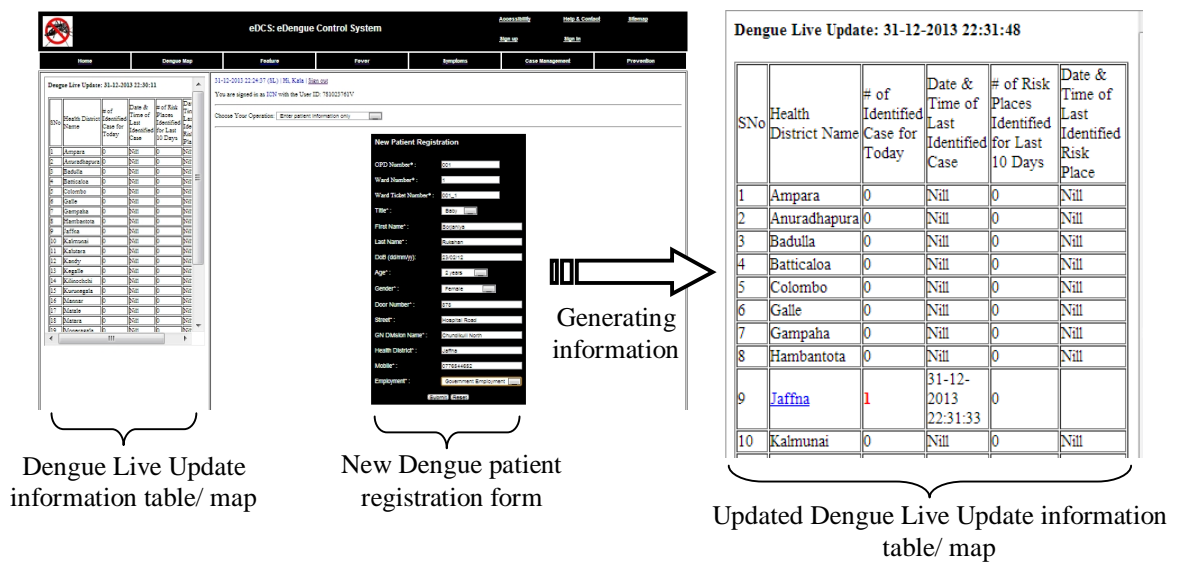

Dengue Live Update information table/ map    New Dengue patient registration form

Generating information

Updated Dengue Live Update information table/ map

**Figure 4: New Dengue patient registration in Health Institutes (Hospitals)**

Dengue Live Update
information table/ map

Detail information for PHI

**Figure 5: The information generated by the system to PHI**

Dengue Live Update
information table/ map

Updated Dengue Live Update information
table/ map

**Figure 6: Patient travel history tracking by PHI and possible risk place generation by the system**

## 5. DISCUSSION

The system has been designed for health professionals to meet their specific needs in Sri Lanka to control the Dengue disease in the country.

The following describes how our system has been developed to overcome the issues which are found in the present system.

**Data Collection**: A data collection module has been developed to collect data effectively for the whole cycle of data collection. Normally, humans are having writing and typing errors. Our system contains predefined data which helps to minimize the human typing errors. Thus, it improves the data collection from citizens, laboratory, hospital, and health care professionals by having computerized online forms. Thus the quality of data is improved by eliminating human typing errors. Furthermore, our system improves the data collection by automating form fields using the suggesting technology, which can be easily written or select during the writing/ typing form. All of above techniques are improving the ability to enforce data input rules on the system which will reduce the invalid input of data thereby increasing the data quality.

**Early Detection:** To identify and manage the dengue epidemic situation, quality and accurate notification is required. The notification alerts will be sent via email and SMSs to take quick action. As our proposed system, collects quality data, and produce accurate information of notification by processing as real time, the health care professionals monitor and take appropriate action instantly for saving lives and protecting peoples, efficiently and timely.

**Information Analysis**: Tailor-made online tool has been developed for data analysis. Automatically generated accurate information is available to provide comparison capability for easy analysis. This facility enables public and private medical consultants to incorporate with the system for effective clinical and case management of the patient. Furthermore, the regional level public health officers can able to analyze the data and monitor their regions, other than the monitoring and evaluation done by the national level.

**Decision Making:** The decision taken will be more accurate and relevant due to the high quality data, accurate information, early case detection, increased capabilities in data analysis. This will give more force to the health care professionals, medical consultants, and others for preventing the dengue disease.

### 5.1 The rapid timely response to control the Dengue outbreak

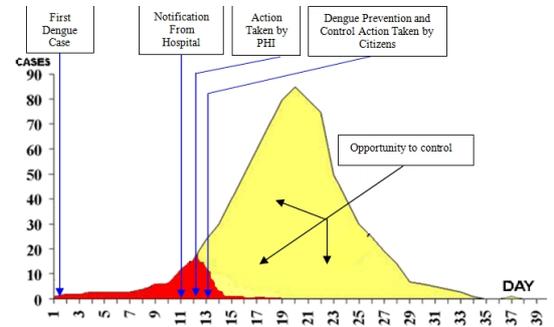

**Figure 7: The rapid timely response to control the Dengue outbreak**

The Figure 7 shows the level of performance delivered by our system to control a Dengue outbreak. Thus, the system only wants two to three days to complete the prevention and control cycle. The rapid detection and timely response to a Dengue outbreak can reduce the magnitude of the outbreak, and yield more opportunity to control the Dengue which reduces the widely spreading dengue disease among the population of Sri Lanka.

Furthermore, at present the timely receiving of weekly return of communicable diseases (H399) for Dengue is approximately 75% (Epidemiology Unit, Ministry of Health, Sri Lanka, 2012). The system will be conveyed to the health care professionals attached to Medical Officers of Health (MOH) for concerning immediate action against the outbreak. The PHI can able to handle the work order and the patient information which is given by the system for attending the case immediately by integrating ICT resources and services, and implement the prevention and control measures as early as possible. The system automates the confirmed cases' report which is to be sent to the Regional Epidemiologist (RE) and the Epidemiologist (E) of the Epidemiology Unit (EU). Furthermore, the report generated data and information will be available in any time for further investigation by the health care professions, technical experts, and researchers. Then, the timely return of weekly return of communicable diseases for Dengue will be 100%. The faster, non repetitive and efficient notification of Dengue will allow the accurate and rapid detection of Dengue communicable diseases and their spreading exact location. Then the citizen can be alert about the disease through the social media which will yield the citizens to actively participate for controlling the dengue disease. The system information is helps to the travellers, and tourist from all over the world to know the disease burden places and which will help to control the human host movement.

## 6. CONCLUTION

Dengue has become hyper endemic in Sri Lanka. The disease has become a huge threat to health of Sri Lankans' and to the social-economic stability as well (Rupasinghe et al, 2010). Dengue fever spread rapidly among people living in most of the districts of Sri Lanka. Furthermore, the Dengue vector mosquitoes are found wide spread in the Southern Province. Now, the political environment in the country is so conductive that number of people travelling from south of Sri Lanka to the districts in Northern Province is increasing. So, if one person among people coming here is incapacitated with dengue, there is tendency towards spread of infection of it. Correct identification in time of these species and breading places are important for dengue and chikungunya control in regional and national level (Weerasinghe et al, 2011).

An integrated ICT-based real-time solution for controlling dengue epidemic has been proposed which is cost effective and fast for the controlling the dengue in regional and country. The web portal system has been designed with minimum implementation and recurrent costs critical for developing countries like Sri Lanka. Then the system has been named as eDCS: eDengue Control System which collects data as early as possible and converting information to early action for preventing and controlling Dengue by integrating ICT resources and services in Sri Lankan prospective. The motto of the eDCS is as follows "Transforming information to early action for serving lives and protecting citizens". Our approach delivers the following outcomes to enhance DF/DHF surveillance through ICT resources and services and the future expansion of surveillance strategy of the system is to include outpatients and community level case finding using suitable mechanisms: It ensures the case according to the definitions for DF/DHF; It support of the laboratory surveillance; It insert of private sector (institutions as well as private practitioners) into the mainstream of the surveillance system; It strengthening of Epidemic Preparedness and response at all levels; It consolidates of notification through managerial and supervisory inputs at all levels of the health system; It provide a rapid, timely inform to Regional Epidemiologist, and the Epidemiologist in the mainstream of the surveillance system; It flow and analysis of data from periphery to centre in a phased manner and the feed back in the reversed direction. The system is initially limited to dengue communicable disease. It can be easily expanded to other communicable diseases, and non communicable disease surveillance in future.